\documentclass[12pt]{article}
\usepackage{amssymb}
\usepackage{amsmath}

\begin{document}

\title{Phase equivalent potentials, Complex coordinates and Supersymmetric Quantum Mechanics}
\author{C.V.Sukumar\\
Department of Physics, University of Oxford\\
Theoretical Physics, 1 Keble Road, Oxford OX1 3NP}
\maketitle

\begin{abstract}
Supersymmetric Quantum Mechanics may be used to construct reflectionless potentials and phase-equivalent potentials. The exactly solvable case of the $\lambda sech^2$ potential is used to show that for certain values of the strength $\lambda$ the phase-equivalent singular potential arising from the elimination of all the boundstates is identical to the original potential evaluated at a point shifted in the complex cordinate space. This equivalence has the consequence that certain general relations valid for reflectionless potentials and phase-equivalent potentials lead to hitherto unknown identities satisfied by the Associated Legendre functions. This exactly solvable probelm is used to demonstrate some aspects of scattering theory. 
\end{abstract}


\vfill\eject

\section{\noindent Introduction}

The connection between muti-soliton solutions of the Korteweg-deVries (KdV) equation and reflectionless potentials in non-relativistic Quantum Mechanics is well known (Scott {\it et al} 1973). For studying the boundstates of quarks with definite angular momentum  solutions of the non-relativistic Schroedinger equation in a confining potential have been considered. The s-state solution for the zero angular momentum case can be viewed as the odd states of a symmetric confining potential in the infinite space $-\infty \le x \le \infty$ (Thacker {\it et al} 1978, Quigg and Rossner 1981). Since confining potentials have no scattering states and have only boundstates the confining potentials belong to the category of symmetric reflectionless potentials. Reflectionless potentials may also be constructed by starting from the free particle case and adding the boundstates by the methods of Supersymmetric Quantum Mechanics (SUSYQM) (Sukumar 1986). The two methods of constructing reflectionless potentials have been shown to be equivalent. 

SUSYQM can also be used to start from a potential with a given spectrum and scattering phaseshifts and eliminate the boundstates to find a potential which is phase equivalent to the starting potential but supporting fewer boundstates (Baye 1987, Sukumar 1985). This construction leads to a definite relation between the two potentials. In this paper we consider the exactly solvable case of the $\lambda sech^2$ potential which corresponds to the category of reflectionless potentials for special values of the strength $\lambda$. We consider this potential in $r$-space and its phase-equivalent partner. Various mathematical properties of the solutions in the two potentials are of interest. In section 2 of this paper we examine the relation between the solutions in two potentials. In section 3 we show that the general theories of relectionless potentials and phase-equivalent potentials lead to certain identities. The procedure is illustrated in section 4 by using a simple example. Section 5 contains a discussion of the main results of the paper. Units in which $\hbar =1$ and the mass $\mu=\frac{1}{2}$ are used throughout this paper so that $\frac{{\hbar}^{2}}{2\mu} =1$. 

\medskip
\noindent
\section{ $Sech^2$ and $Cosech^2$ Potentials and Complex Coordinates}

We consider the Schroedinger equations for two potentials, one attractive and the other repulsive, of the form 
\begin{align}
V_{d}\ &=\ -2n(2n+1)\ {\cosh}^{-2}r \\
V_{s}\ &=\ +2n(2n+1)\ {\sinh}^{-2}r \label{}
\end{align}
for integer values of $n$ and energy $E_j =-\gamma_j^2$ so that
\begin{align}
\frac{d^2 \Psi_{j}}{dr^2}\ &=\ \left(-2n\left(2n+1\right)\cosh^{-2} r\ +
\ \gamma_j^2\right) \Psi_j \ ,\\
\frac{d^2 \Phi_{j}}{dr^2}\ &=\ \left(+ 2n\left(2n+1\right)\sinh^{-2} r\
+\ \gamma_j^2\right) \Phi_j \ .\label{}
\end{align}
Under the substitution $z=\tanh r$ eq. (3) transforms to
\begin{equation}
\left[\frac{d}{dz} \left(1 - z^2\right) \frac{d}{dz}\ +\
2n\left(2n+1\right)\ -\ \frac{{\gamma}_{j}^{2}}{1-z^2}\right] \Psi_j\
=\ 0 \label{}
\end{equation}
which is the differential equation satisfied by the Associated Legendre functions. Polynomial solutions arise for integer values of $\gamma_{j}$. The $n$ boundstate energy eigenvalues of $V_d$ and the corresponding normalised eigenfunctions satisfying vanishing boundary conditions at $z=0$ and $z=1$ are given by
\begin{align}
\gamma_ j\ &=(2j-1)\ ,\ E_j\ =\ -{\gamma}_{j}^{2} \ ,\ \ j=1,2,..,n \notag\\
 \Psi_{j} &= \alpha_j  P_{2n}^{2j-1}\left(\tanh
r\right)\ ,\ \alpha_{j}= {\sqrt{\frac{2(2j-1)\left((2n+1-2j)!\right)}{(2n-1+2j)!}}}
 \label{}
\end{align}
with $j=n$ corresponding to the groundstate and $j=1$ to the highest lying boundstate. The associated Legendre functions can be given in the explicit form (Abaramowitz and Stegun 1965)
\begin{align}
z<1\ ,\ P_n^m(z)\ &=\ (-)^m \frac{(1-z^2)^{\frac{m}{2}}}{2^n\ n!}
\ \frac{d^{n+m}}{dz^{n+m}}\ \left(1-z^2\right)^n \ ,\notag \\
y>1\ ,\ P_n^m(y)\ &=\ +\ \ \frac{(y^2-1)^{\frac{m}{2}}}{2^n\ n!}
\ \frac{d^{n+m}}{dy^{n+m}}\ \left(y^2-1\right)^n \ . \label{}
\end{align}

\bigskip

$V_s$ is a repulsive potential which supports no boundstates. However it is still possible to find a solution to eq. (4) at energies corresponding to the boundstates of $V_d$. Under the substitution $y=\coth r$ eq. (4) transforms to the same form as eq. (5) with $y$ replacing $z$ and therefore the solutions in $V_s$ at energies corresponding to the boundstates of $V_d$ may be given in the form
\begin{equation}
\gamma_j=(2n+1-2j)\ ,\ \Phi_{j} \ =\ - \alpha_j \ P_{2n}^{2n+1-2j} \left(\coth
r\right) \ ,\ j=1,2..n\ . \label{}
\end{equation}
The additional minus sign in the expression for $\Phi_j$ is placed to ensure that $\Phi_j$ and $\Psi_j$ have the same asymptotic limit as $r\to\infty$.

\bigskip

Under the coordinate transformation $r\to r+i\pi/2$
\begin{align}
\sinh r \to i\cosh r\ ,\  \cosh r &\to i\sinh r\ ,\ \tanh r \to \coth r
\ ,\notag\\
{\cosh}^{-2} r &\to -{\sinh}^{-2} r\ ,\   V_d \to V_s \label{}
\end{align}
which enables the identification
\begin{equation}
\Phi_j \left(r\right)\ \sim \ \Psi_j \left(r + i \frac{\pi}{2}\right)\ .
\label{}
\end{equation}
In fact the two functions in eq. (10) are equal within a possible phase factor which may be chosen so that the unnormalisable functions $\Phi_j(r)$ have the same asymptotic limit as $\Psi_j(r+i\pi/2)$ when $r\to\infty$. 

\bigskip
\noindent

Thus we have shown that the eigenfunctions at the same energy for the two potentials are simply related by a shift of the coordinate in the complex $r$-plane. We have considered the solutions in these special potentials because they play an active role in the discussion in the next section.   

\bigskip

\section{\protect\bigskip Reflectionless Potentials and Phase-equivalent Potentials}

A symmetric reflectionless potential in $-\infty \leq x \leq +\infty$ with binding energies $E_j =-{\gamma}_{j}^{2}, j=1,2,...N$ may be constructed by adding boundstates to a free particle potential $V_0=0$ using a method for addition of boundstates based on Supersymmetric
Quantum Mechanics (Sukumar 1986) and is given by
\begin{align}
V(x)\ &=\ -2 \frac{d^2}{dx^2}\ \ln Det M \notag \\ 
M_{kj}(x)\ &=\ \frac{{\gamma}_{j}^{k-1}}{2}\ \left(\exp \gamma_j x +
\ \left(-1\right)^{j+k} \exp -\gamma_j x\right)\ ,\ k,j=1,2,...,N .\label{}
\end{align}
It has been shown that it is possible to represent $V$ in terms of the normalised eigenstates in the form
\begin{equation}
V(x)\  = \ -4\ \sum_{j=1}^{N} \gamma_j {\Psi}_{j}^{2}\ . \label{}
\end{equation}
This potential is equivalent to the $N$ soliton solution of the KdV at time $t=0$ constructed by other methods ( Scott {\it et al} 1973). If the $\gamma_j$ are chosen to be integers in the sequence [1,2,..,N] then the the potential defined by eq.(11) can be shown to be a $sech^2x$ potential (see Appendix for a proof), ({\it i.e}):
\begin{equation}
V(x)\ =\ -2\ \frac{d^2}{dx^2}\ln DetM\ =\ -\ \frac{N(N+1)}{\cosh^2 x}\
.\label{}
\end{equation}
The eigenstates of this potential are Associated Legendre polynomials as shown by eq.(6) in the last section. The representation of $V$ in terms of normalised eigenstates in eq. (11) and the variable $z=\tanh x$ may be used to obtain the relation
\begin{equation}
N(N+1)\ (1-z^2)\ =\ 4\ \sum_{m=1}^{N} m^2\ \frac{(N-m)!}{(N+m)!}\
\left(P_N^m\left(z\right)\right)^2 \ .\label{}
\end{equation}
The equality identified in eq. (14) seems to be a new result and is not listed in the usual texts on the properties of Associated Legendre polynomials.

\bigskip

If $N$ is chosen to be an even number $N=2n$, the antisymmetric states of $V(x)$ may be viewed as the $n$ eigenstates of a potential $V(r),\ 0\leq r\leq \infty$. We now concentrate on the radial domain and choose $N=2n$. If the $\gamma_j$ values in eq. (11) are in the integer sequence [1,2..,2n] the potential arising from eq.(11), after the change of variable $x\to r$, is the potential $V_d$ defined in eq. (1). As noted before the $n$ boundstates of $V_d$ correspond to $\gamma_j$ values in the odd integer sequence [1,3,..,2n-1]. We next examine the scattering states of $V_d$. Using the free particle solution $\sin\kappa r$ at energy $E=\kappa^2$ and a general result connecting the eigenfunctions of supersymmetric partner
potentials at the same energy (see Appendix) it is possible to represent the scattering state of $V(r)$ for the same positive energy $E$ in the form
\begin{equation}
\Psi(\kappa,r)\ =\ \frac{Det D(r)}{Det M(r)} \label{}
\end{equation}
where the matrix $M$ is of dimension (NxN) and the elements of the matrices $M$ and $D$ are given by
\begin{align}
D_{kj}(r)\ &=\ M_{kj}(r)\ , \ k,j=1,2...,2n \notag \\
D_{2n+1,j}(r)\ &=\ M_{2n+1,j}(r)\ , \ j=1,2..,2n \notag \\
M_{kj}(r)\ &=\ \frac{j^{k-1}}{2} \left( \exp jr +
(-)^{j+k}\exp-jr \right) \notag \\
D_{k,2n+1}(r)\ &= \frac{d^{k-1}}{dr^{k-1}} \sin \kappa r\ ,\ k=1,2...,2n+1.
\label{}
\end{align}
The elements of the matrices $M$ and $D$ have a simple form in the limit $r\to \infty$ because the exponentially decaying parts of $M_{kj}$ vanish in this limit. The exponentially growing parts scale and contribute a term to the determinant of M which cancels a corrsponding term coming from the determinant of D. It may be shown that if a matrix $A$ has elements $A_{kj}={\gamma}_{j}^{k-1},\ j,k=1,2,..,N$ then 
\begin{equation}
Det A\ =\ \prod_{j=1}^{N-1}\ \prod_{k>j}^{N}\ \left(\gamma_k
\ -\ \gamma_j\right) .\label{}
\end{equation} 
Using this theorem repeatedly by setting $\gamma_{2n+1}=\pm i\kappa$ the determinants of the matrices $M$ and $D$ may be evaluated in the $r\to \infty$ limit to give
\begin{equation}
Lt_{r\to \infty}\ \Psi(\kappa,r) = -\frac{i}{2} \left(\exp i\kappa r\right)
\prod_{j=1}^{2n}\left(-j +i\kappa\right) + \frac{i}{2} \left(\exp
-i\kappa r\right) \prod_{j=1}^{2n}\left(-j -i\kappa\right)\label{}
\end{equation}
and hence the phaseshift may be calculated to be 
\begin{equation}
\delta = \frac{1}{2i} \ln\left(\prod_{j=1}^{2n}\frac{+i\kappa -
j}{-i\kappa - j}\right) = 
 n\pi - 
\sum_{j=1}^{2n}\arctan \frac{\kappa}{j} \
.\label{}
\end{equation}
It is clear that this phaseshift relation satisfies Levinson's theorem (Levinson 1949, Swan 1968) which states that the phaseshift at zero energy must equal $\pi$ multiplied by the number of boundstates. 

\bigskip
We next study the properties of the repulsive potential $V_s(r)$ in eq. (2). Since $V_s$ can be obtained from $V_d$ by the mapping $r\to r+i\pi/2$ eqs. (11) and (12) may be used to represent $V_s$ in the form 
\begin{align}
V_s(r) &= -2 \frac{d^2}{dr^2}\ \ln Det {\tilde M}\ , \notag\\
{\tilde M}_{kj}(r) &= M_{kj}\left(r+i\frac{\pi}{2}\right)\ ,\notag\\
M_{kj}(r) &= \frac{j^{k-1}}{2}\left(\exp jr +
(-)^{j+k} \exp -jr\right),\ k,j=1,2,..,2n . \label{}
\end{align}
The scattering states of $V_s$ for $E=\kappa^2$ may be given in the form
\begin{align}
\Phi(\kappa,r)\ &=\ \frac{Det{\tilde D}(r)}{Det{\tilde M}(r)} \
,\notag\\
{\tilde D}_{kj}(r)\ &=\ {\tilde M}_{kj}(r)\ ,\ k,j=1,2,..,2n \notag\\
{\tilde D}_{2n+1,j}(r)\ &=\ {\tilde M}_{2n+1,j}(r)\ , \ j=1,2..,2n
\notag{}\\
{\tilde D}_{k,2n+1}(r)\ &=\ \frac{d^{k-1}}{dr^{k-1}} \sin \kappa r \ ,\
k=1,2,..,2n+1 \ .\label{}
\end{align} 
The same reasoning as that used to go from eq. (15) to eq. (18) for deriving the asymptotic form of $\Psi(\kappa,r)$ may be used to find the asymptotic form of $\Phi(\kappa,r)$ in the limit $r\to\infty$. The additional factors arising in the evaluation of the determinants of ${\tilde M}$ and ${\tilde D}$ due to the transformation $r\to r+i\pi/2$ are the same in the limit $r\to\infty$ and cancel each other. Hence it is possible to show that
\begin{equation}
Lt_{r\to\infty} \ \Phi(\kappa,r)\ =\ Lt_{r\to\infty}\ \Psi(\kappa,r)
\label{}
\end{equation}
leading to the result that $V_d$ and $V_s$ defined by eqs. (1) and (2) have phaseshifts which are equal within integral multiples of $\pi$ for all positive energies. $V_d$ supports $n$
boundstates but $V_s$ is a repulsive potential with no boundstates. $V_s$ has a repulsive singularity at the origin of the form $2n(2n+1)/r^2$.

\bigskip

It is known that when the phaseshifts are equal within integral multiples of $\pi$  for all positive energies but the boundstates are missing then the deep potential with boundstates and
the singular potential without any boundstates may be related  by supersymmetry and the difference between the potentials can be expressed as (Baye 1987, Baye and Sparenberg 1994, Sukumar and Brink 2004)
\begin{align}
V_s\ &=\ V_d\ -\ 2 \frac{d^2}{dr^2} \ln Det F \notag\\
F_{jk}\ &=\ \int_{0}^{r} \Psi_j (y)\ \Psi_k (y) dy \ .\label{}
\end{align}
The solutions in $V_s$ at energies corresponding to the missing boundstates of $V_d$ may be found by solving
\begin{equation}
F_{jk}\ \Phi_{k}(r)\ =\ \Psi_j (r)\  ,\ \ Lt_{r\to\infty}\ \Phi_j (r)\  =
\ Lt_{r\to\infty}\ \Psi_j (r)\ . \label{}
\end{equation}
It may also be shown that in terms of the solutions in the two potentials at the energies
corresponding to the boundstate energies of the deep potential 
\begin{align}
\frac{d}{dr} \ln Det F\ &=\ \sum_{j=1}^{n} \Psi_j(r)\ \Phi_j(r)\ ,
\notag\\
V_s\ -\ V_d\ &=\ -2 \frac{d}{dr}\ \sum_{j=1}^{n} \Psi_j(r)\ \Phi_j(r) \ .\label{}
\end{align}
Using the explicit forms of $V_d$ and $V_s$ given in eqs. (1) and (2) eq. (25) may be integrated
from $r$ to $\infty$ to give
\begin{equation}
\frac{n(2n+1)}{\sinh r\ \cosh r}\ =\ \sum_{j=1}^{n}\Psi_j(r)\ \Phi_j(r)
\ .\label{}
\end{equation}
Using the expressions for the wavefunctions given in eqs. (6) and (8) and the variable $z=\tanh r$ we can establish the equality
\begin{equation}
 -\ \sum_{m=1,3,..}^{2n-1} P_{2n}^{m}\left(z\right)\ P_{2n}^{m}\left(\frac{1}{z}\right)\ 2m
\ \frac{(2n-m)!}{(2n+m)!}\ = \ n(2n+1)\ \frac{1-z^2}{z} .\label{}
\end{equation}
This is another new identity not listed in the usual texts on the properties of Associated Legendre functions.

\bigskip
\noindent
\section{A Simple Example}
The results in eqs. (14) and (26) may be illustrated by considering the case $N=2, n=1$. Using
\begin{equation}
P_2^1(z)\ =\ -3z\sqrt{1-z^2}\ ,\ P_2^2(z)\ =\ 3(1-z^2) \label{}
\end{equation}
it can be checked that
\begin{equation}
6(1-z^2)\ =\ 4\left( 1^2\ \frac{1!}{3!} \left(P_2^1(z)\right)^2\ +\
2^2\ \frac{0!}{4!} \left(P_2^2(z)\right)^2\right) \label{}
\end{equation}
verifying the equality in eq. (14) for the case $N=2$.

\bigskip

Using
\begin{align}
\Psi_1(r)\ &=\ -\sqrt{3}\ \frac{\tanh r}{\cosh r}\ ,\ F_{11}\ =\int_0^r
\Psi_1^2(y) dy\ =\ \tanh^3 r \notag \\
F_{11}\ \Phi_1\ &=\ \Psi_1\ ,\ \Phi_1(r)\ =\ -\sqrt{3}\ \frac{\coth
r}{\sinh r} = i \Psi_1 (r+i\frac{\pi}{2})\label{}
\end{align}
it can be seen that
\begin{equation}
\Psi_1(r)\ \Phi_1(r)\ =\ +\frac{3}{\sinh r\ \cosh r}  \label{}
\end{equation}
thereby verifying eq. (26) for the case $n=1$.

\bigskip

The scattering state of the potential in eq. (1) for the case $n=1$ for positive energy $E=-\kappa^2$ constructed using eqs. (12), (15) and (16) with $\gamma_1=1$ and $\gamma_2=2$ is given by
\begin{equation}
\Psi(\kappa,r)\ =\ \left(\kappa^2 -2 + \frac{3}{\cosh^2 r}\right)\ \sin \kappa
r\ +\ 3 \kappa \tanh r \ \cos\kappa r \label{}
\end{equation}
leading to the phaseshift
\begin{equation}
\delta\ =\ \pi\ -\ \arctan \kappa \ - \ \arctan \frac{\kappa}{2} \
.\label{}
\end{equation}
The scattering state of the potential in eq. (2) for the case $n=1$ constructed using eqs. (20) and (21) is given by
\begin{equation}
\Phi(\kappa,r)\ =\ \left(\kappa^2 - 2 - \frac{3}{\sinh^2 r}\right)\ \sin
\kappa r\ + \ 3\kappa \coth r \ \cos \kappa r \label{}
\end{equation}
leading to the phaseshift
\begin{equation}
{\tilde \delta}\ =\ -\arctan \kappa \ -\ \arctan \frac{\kappa}{2} _
.\label{}
\end{equation} 
It was noted earlier that $V_s$ is the phase-equivalent singular potential arising from $V_d$ by the elimination of the boundstates of $V_d$. Eqs. (33) and (35) show that the zero energy phaseshifts satisfy Levinson's theorem (Swan 1968) as applied to the case of a missing boundstate and that for all positive energies the phaseshifts differ by $\pi$. Thus the phaseshifts of the two potentials are equal within a multiple of $\pi$ when $n=1$.

\bigskip

From the general theory for the construction of singular potentials by the elimination of boundstates it may be shown that the wavefunctions in the phase-equivalent deep and singular potentials are related by
\begin{equation}
\Phi(\kappa,r)\ =\ \Psi(\kappa,r)\ -\ \sum_{j=1}^{n}\Phi_j (r)\
\int_{0}^r \Psi_j (y) \Psi(\kappa,y) dy \label{}
\end{equation}
which may be simplified by using the Wronskian between $\Psi_j(r)$ and $\Psi(\kappa,r)$ to the form
\begin{equation}
\Phi(\kappa,r)\ =\ \Psi(\kappa,r)\ +\ \sum_{j=1}^{n}\frac{\Phi_j
(r)}{\gamma_j^2 +\kappa^2} \ \left(\Psi_j(r)\frac{d}{dr}\Psi(\kappa,r)
- \Psi(\kappa,r)\frac{d}{dr}\Psi_j(r)\right) \ . \label{}
\end{equation}
It may be verified that eqs. (32) and (34) satisfy eq. (37).

\bigskip

In scattering theory the phaseshift may be expressed in terms of the scattering wavefunction normalised to the asymptotic form
$Lt_{r\to\infty}\Psi(\kappa,r)=\sin(\kappa r+\delta)$ by the exact expression (Messiah 1958)
\begin{equation}
\sin \delta\ =\ - \frac{1}{k} \int_0^{\infty} V(r) \sin \kappa r\
\Psi(\kappa,r) dr \ .\label{}
\end{equation}
For the case $n=1$ use of the scattering wavefunction in eq. (32) with appropriate changes to account for the different normalisation leads to the integral relation
\begin{equation}
1 = +\frac{2}{\kappa^2} \int_0^{\infty} \frac{\sin\kappa r}{\cosh^2
r}\left(\left(\kappa^2 -2 +\frac{3}{\cosh^2r}\right)\sin\kappa r + 3\kappa
\tanh r \cos\kappa r\right)
\end{equation}
which can be verified by explicit evaluation of the integral.

\bigskip

The Born approximation limit of the phaseshift for the potential $V_d = -6 \cosh^{-2}r$ obtained by replacing $\Psi(\kappa,r)$ by $\sin \kappa r$ in eq. (38) is
\begin{equation}
\sin\delta \ =\ \frac{6}{\kappa}\int_0^{\infty}
\frac{\sin^2 \kappa r}{\cosh^2 r}
 \ dr\ =\
\frac{3}{k}\left(1-\frac{\kappa\pi}{\sinh\kappa\pi}\right) \label{}
\end{equation}
leading to
\begin{equation}
Lt_{\kappa\to\infty}\ \sin \delta\ =\ \frac{3}{\kappa},\ \ 
Lt_{\kappa\to\infty}\ \delta\ =\ \frac{3}{\kappa} \label{}
\end{equation}
which is in agreement with the limiting value of the phaseshift obtained from eq. (33) as given by
\begin{equation}
Lt_{\kappa\to\infty}\ \delta\ =\ \pi\ -\ \left(\frac{\pi}{2} -
\frac{1}{\kappa}\right)\ -\ \left(\frac{\pi}{2} -
\frac{2}{\kappa}\right)\ =\ \frac{3}{\kappa}\ .\label{}
\end{equation}

\bigskip

\section{Discussion}

In this paper we have considered attractive $sech^2$ potentials of strength $\lambda =2n(2n+1)$ for integer values of $n$. In the infinite space $-\infty \le x \le \infty$ this is a symmetric reflectionless potential. In the semi-infinite space $0 \le r \le \infty$ this potential has a definite phase-shift for positive energies. By eliminating all the boundstates of this potential it is possible to find a phase-equivalent potential which is a singular repulsive $cosech^2$ potential of strength $\lambda=2n(2n+1)$. The eigenstates of these two potentials have been shown to be related by a shift of coordinate in the complex plane. If the strength $\lambda \ne 2n(2n+1)$ then the potential in the $x$-space is not of the reflectionless category. Furthermore the phase equivalent singular potential generated by eliminating the boundstates by the SUSY procedure is not simple and is not of the form of a $cosech^2$ potential. Hence $sech^2$ potential of strength $\lambda =2n(2n+1)$ belongs to a special category and we have shown that the solutions in this potential and its phase-equivalent partner lead to the identities given in eqs. (14) and (27). We have shown that the exactly solvable example considered in this paper enables a direct demonstration of the exactness of the general expression for the phase shift (eq. (38)) in scattering theory. The relation between potentials, their phase-equivalent partners and transformations in complex coordinate space is an interesting subject which merits further study. Eq. (12) which expresses the reflectionless potentials in terms of their eigenstates and eq. (25) which expresses the difference between a potential with boundstates and its phase-equivalent partner with fewer boundstates in terms of the solutions in the two potentials at the energies of the eliminated boundstates encapsulate a nonlinear structure which is quite general and not restricted to simple exactly solvable models.

\section{\protect\bigskip Acknowledgement}
I thank a referee for pointing out that eq. (14) is a special
case of an identity listed in the book Integrals and Series (volume
3) authored by Prudnikov, Brychkov and Marichov (1990).

\section{\protect\bigskip Appendix}

Supersymmetric Quantum Mechanics (Sukumar 1985)) may be used to remove or add boundstates to a given potential. If we start from the potential
\begin{equation}
V_N(x)\ =\ - \frac{N(N+1)}{\cosh^2 x} \label{}
\end{equation}
then it may be verified that
\begin{equation}
\xi_N(x)\ \sim \ \cosh^{-N} x \label{}
\end{equation} solves the Schroedinger equation for the potential $V_N$ with the energy eigenvalue $E_N=-N^2$. Since $\xi_N$ is a nodeless nomalisable function it must be the groundstate of $V_N$. The normalised groundstate eigenfunction normalised to unity in the interval $[-\infty,+\infty]$ (Gradshteyn and Ryzhik 1965) is
\begin{equation}
\Psi_N(x)\ =\ \alpha_N\ \cosh^{-N}x\ ,\ \alpha_N\ =\
\sqrt{\frac{\left(2N-1\right)!!}{2^{N}\left(N-1\right)!}} .\label{}
\end{equation}
The supersymmetric partner to $V_N$ which has the same spectrum as $V_N$ except for missing the groundstate at $E_N$ is given by
\begin{equation}
{\tilde V}_{N}(x)\ =\ V_N(x)\ -\ 2\ \frac{d^2}{dx^2}\ln \xi_N(x) \ =\ -
\frac{(N-1)N}{\cosh^2 x} \ . \label{}
\end{equation}
The normalised eigenfunctions $\Psi$ of $V_N$ and the normalised eigenfunctions $\Phi$ of
${\tilde V_N}$ for all energies $E\ne E_N$ are related by
\begin{align}
\Phi(E,x) \ &= \ \frac{1}{\sqrt{(E-E_N)}}\ A_N^- \Psi(E,x) \ ,\ \ \Psi(E,x) \ =
\ \frac{1}{\sqrt{(E-E_N)}}\ A_N^+ \Phi(E,x) \ \notag\\
A_N^{\pm} \ &= \ \pm \frac{d}{dx}\ +\
\frac{1}{\xi_N}\frac{d\xi_N}{dx}\ .\label{}
\end{align}
These intertwining relations are also valid for the scattering states for energy $E=\kappa^2$ and may also be given in the form
\begin{align}
\Phi(\kappa,x)\ &= \frac{det D(x)}{\Psi_N(x)} \notag\\
D_{11}&= \Psi_N(x),\ D_{12}=\Psi(\kappa,x),\
D_{21}=\frac{d}{dx}\Psi_N(x),\ D_{22}=\frac{d}{dx}\Psi(\kappa,x)\
.\label{}
\end{align}It is clear from eqs. (43) and (46) that ${\tilde V}_N$ may be obtained from $V_N$ by the transformation $N\to (N-1)$. Hence it may be concluded that the groundstate energy of $V_{N-1}$ is $E_{N-1}=-(N-1)^2$ with the groundstate eigenfunction
\begin{equation} 
\xi_{N-1}(x)\ \sim\ \cosh^{-(N-1)}(x) \ .\label{}
\end{equation}

Using the intertwining operator $A_N^+$ in eq. (47) the first excited state of $V_N$ may be found in the form
\begin{equation}
\Psi_{N-1}\ \sim \ \frac{1}{\Psi_N}\ \frac{d}{dx}\ \left(\Psi_N
\xi_{N-1}\right)\ \sim\ \cosh^Nx\ \left(\frac{d}{dx}
\frac{1}{\cosh x}\right)\ \cosh^{2-2N}x \ .\label{}
\end{equation}
This procedure may be iterated $N$ times to show that the eigenvalue spectrum of $V_N$ is given by $E_j= - j^2 , j=1,2,..,N$ and that the potential after $N$ iterations is
\begin{align}
V_0\ &=\ V_N\ -\ 2 \frac{d^2}{dx^2} \ln \xi \notag\\
\xi\ &=\ \prod_{j=1}^{N} \xi_j \ =\ \left(\cosh x\right)^{-\frac{N(N+1)}{2}}
\end{align}
which when evaluated gives $V_0=0$. By iteration of eq. (50) the eigenfunctions $\Psi_{N-j}$ of
$V_N$ with eigenvalues $E_{N-j}=-(N-j)^2, j=0,1,2,..,N-1$ may be given in the form
\begin{align}
\Psi_{N-j}\ &\sim\ \frac{1}{\xi_N}\ A_N^{+} A_{N-1}^{+} ... A_{N+1-j}^{+} \
\xi_{N-j} \notag\\
\ &\sim\ \cosh^N x\ \left(\frac{d}{dx}\frac{1}{\cosh x}\right)^j\
\cosh^{2j-2N}x \ ,\ j=0,1,..,N-1.\label{}
\end{align}
The eigenstates may be normalised by including the energy denominators arising from factors similar to the ones appearing in eq. (47) and the normalisation factor $\alpha_{N-j}$ which can be found using eq. (45). The normalised eigenstates are
\begin{equation}
\Psi_{N-j}=(2N-2j-1)!!\sqrt{\frac{(N-j)}{(2N-j)!j!}}\cosh^Nx\left(\frac{d}{dx}\frac{1}{\cosh
x}\right)^J \cosh^{2j-2N}x . \label{}
\end{equation}
Since the Schroedinger equation for $V_N$ may be shown to be identical to the differential equation for the Associated Legendre functions in terms of the variable $z=\tanh x$ the eigenfunctions are Associated Legendre polynomials in $z$ and the normalised eigenstates may be
given in the form
\begin{equation}
\Psi_{N-j}(x)\ =\ \sqrt{\frac{j!}{(2N-j)!}(N-j)}\ P_N^{N-j}(\tanh x)\ ,\ j=0,1,2,..N-1 \ .\label{}
\end{equation}
Comparison of eqs. (53) and (54) provides a new representation of the Associated Legendre polynomials given by
\begin{equation}
P_N^{N-j}(\tanh x) = \frac{(2N-2j-1)!!}{j!}
\cosh^Nx\left(\frac{d}{dx}\frac{1}{\cosh x}\right)^J \cosh^{2j-2N}x
.\label{}
\end{equation}

\bigskip

The above procedure may be reversed to start from the free particle potential $V_0=0$ and SUSYQM may be used to add the boundstates at $E_j=-j^2 , j=1,2,..,N$. It has been shown (Sukumar 1986) that
\begin{align}
V_N\ &=\ V_0\ -\ 2\ \frac{d^2}{dx^2} \ln Det M \notag\\
M_{kj}(x)\ &=\ \frac{{\gamma}_{j}^{k-1}}{2} \left(\exp \gamma_j x\ +\
\left(-1\right)^{j+k} \exp -\gamma_j x\right)\ , k,j=1,2,..,N\notag\\
\gamma_j\ &=\ j \ .\label{}
\end{align}
Comparison of the two expressions relating the potentials viewed as addition of $N$ boundstates or the removal of $N$ boundstates leads to the relation
\begin{equation}
\ln Det M\ =\ -\ln \xi\ +\ \alpha x\ + \ \beta \label{}
\end{equation}
where $\alpha$ and $\beta$ are constants which may be determined by examining the expressions in the limit $x\to\infty$:
\begin{align}
Lt_{x\to\infty}\ \xi\ &=\ 2^{\frac{N(N+1)}{2}}\ \exp -\frac{N(N+1)}{2}
x\ \notag\\
Lt_{x\to\infty}\ Det M\ &=\ 2^{-N}\ \exp \frac{N(N+1)}{2} x\ Det A
\notag\\
A_{kj}\ &=\ j^{k-1}\ ,\ k,j=1,2,..,N \ .\label{}
\end{align}
The determinant of the matrix $A$ whose elements are powers of integers may be evaluated using eq. (17) and by comparison of the two limits it can be shown that
\begin{equation}
\alpha\ =\ 0 \ ,\ \exp \beta\ =\ 2^{\frac{N(N-1)}{2}}\ \prod_{j=1}^{N-1}
j!\ .\label{}
\end{equation}
Thus it can be established that
\begin{equation}
Det\ M \ =\ 2^{\frac{N(N-1)}{2}}\ \left(\cosh
x\right)^{\frac{N(N+1)}{2}}\ \prod_{j=1}^{N-1} j! \ \label{}
\end{equation}
and the resulting potential after the addition of $N$ boundstates at
$E_j=-j^2, j=1,2,..,N$, is
\begin{equation}
V_N\ =\ -2\ \frac{d^2}{dx^2} \ln Det M \ =\ -\frac{N(N+1)}{\cosh^2 x}\
.\label{}
\end{equation}
By generalising the intertwining relations in eqs. (47) and (48) for the case of addition of $N$ boundstates it may be established that the scattering states of $V_N$ for energy $E=\kappa^2$ may be related to the scattering solutions $\Phi(\kappa,x)$ in $V_0$ by 
\begin{align}
\Psi(\kappa,x)\ &=\ \frac{Det D (x)}{Det M (x)} \notag\\
D_{kj}\ &=\ M_{kj}(x)\ ,\ k,j=1,2,..,N \notag\\
D_{N+1,j}\ &=\ M_{N+1,j}(x)\ ,\ j=1,2,..,N \notag\\
D_{k,N+1}\ &=\ \left(\frac{d}{dx}\right)^{k-1}\Phi(\kappa,x)\ ,\
k=1,2,..,N+1\label{}
\end{align}

\bigskip

\section{\protect\bigskip References}

1. Scott A.C., Chu F.Y.E and Mclaughlin D.W. 1973 {\it Proc. I.E.E.E.} {\bf 61} 1443.

\noindent
2. Thacker H.B., Quigg C. and Rosner J.L. 1978 {\it Phys. Rev.} {\bf D18} 274,287.

\noindent
3. Quigg C. and Rosner J.L. 1981 {\it Phys. Rev.} {bf D23} 2625.

\noindent
4. Sukumar C.V. 1986 {\it J. Phys. A: Math. Gen.} {\bf 19} 2297

\noindent
5. Baye D. 1987 {\it Phys. Rev. Let} {\bf 58} 2738.

\noindent
6. Baye D. 1987 {\it J. Phys. A: Math. Gen.} {\bf 20} 5529.

\noindent
7. Sukumar C.V. 1985 {\it J. Phys. A: Math. Gen.} {\bf 18} 2917, 2937.

\noindent
8. Abramowitz M. and Stegun I.A. 1965 {\it Handbook of Mathematical Functions} (New York: Dover) 332.

\noindent
9. Levinson N. 1949 {\it Kgl. Danske Videnskab. Selskab, Mat.-Fys. Medd} {\bf 25} 9.

\noindent
10.Swan P. 1968 {\it Ann. Phys.} {\bf 48} 455.

\noindent
11.Baye D. and Sparenberg J.M. 1994 {\it Phys. Rev. Let.} {\bf 73} 2789.

\noindent
12.C.V.Sukumar and D.M.Brink 2004 {\it J. Math. Phys. A: Math. Gen.} {\bf 37} 5689.

\noindent
13.A.Messiah 1958 {\it Quantum Mechanics} (New York: John Wiley \& Sons, INC) 405.

\noindent
13.Gradshteyn I.S. and Ryzhik I.M. 1965 {\it Table of Integrals, Series and Products} (New York: Academic) 344.

\noindent
14.Prudnikov A.P, Brychkov Yu.A and Marichov O.I 1990 {\it Integrals
 and Series Volume 3} (Gordon and Breach Science Publishers) 387.

\end{document}